\documentstyle[prl,aps]{revtex}
\input epsf

\tighten

\begin{document}

\newcommand{\be}{\begin{equation}} 
\newcommand{\ee}{\end{equation}} 
\newcommand{\ba}{\begin{array}} 
\newcommand{\ea}{\end{array}}

\title{Fermionic Zero Modes on Domain Walls}

\author{Dejan Stojkovic}

\address{
Department of Physics,
Case Western Reserve University,
10900 Euclid Avenue, Cleveland, OH 44106-7079. USA
}

\wideabs{
\maketitle

\begin{abstract}
\widetext
We study fermionic zero modes in the domain wall background. The
fermions
have Dirac and left- and right-handed Majorana mass terms. The source of 
the Dirac mass term is the coupling to the
scalar field $\Phi$. The source of the Majorana mass terms could 
also be the coupling to the scalar field $\Phi$ or the vacuum expectation
value of some 
other field acquired in a phase transition well above the phase transition of 
the field $\Phi$. We derive the fermionic equations of motion  
and find the necessary and sufficient conditions for a 
zero mode to exist. We also find the solutions numerically. In the absence 
of the Majorana mass terms, the equations are solvable analytically. In the case of 
massless fermions a zero energy solution exists and we show that although
this mode is not discretely normalizable it is Dirac delta function normalizable
and should be viewed as part of a continuum spectrum rather than as an
isolated zero mode.

\end{abstract}

}

\narrowtext

\section{Introduction}

The vacuum structure in theories with spontaneous symmetry breaking is very 
rich. There exists topologically stabile configurations of gauge and Higgs 
fields known as monopoles, strings and domain walls. Also, classical 
configurations which are not topologically
stable --- non-topological solitons --- could exist and be stable because
of dynamical reasons. Such objects may play an important role in 
the evolution of our universe. 

Extensive research has been done on the interaction of fermions with
solitons and many new and interesting phenomena have been discovered.
For example fermionic zero modes on strings are responsible for string
superconductivity \cite{Witten}. Quark and leptonic zero modes have an
important effect on the stability of the non-topological electroweak
strings \cite{Naculich,Liu,Groves}. Some work has been done also on fermionic
zero modes on domain walls, but only Dirac fermions have been considered
\cite{Jackiw,Callan}. 
 
Recent experimental results \cite{Kamiokande} strongly suggest that
neutrinos, although very light, have a finite mass. Being
neutral, neutrinos can have 
Majorana masses in addition to the usual Dirac mass terms. This paper 
study zero modes of a particle with 
both Dirac and Majorana mass terms in the background of a domain wall.

In section 2 we present the Lagrangian of a theory where the coupling
to the real scalar field $\Phi$ is the source of all mass terms. We 
derive the equations of motion.
In section 3 we solve the equations analytically in two asymptotic regimes,
near the origin and far from the origin, and find the necessary and 
sufficient condition for the zero mode to exist. We also solve the 
equations numerically. If the Majorana mass terms are absent it is possible
to solve the equations analytically and our result agrees with the
results in literature  \cite{Jackiw,Callan}. We discuss the zero energy
solutions in the case of massless fermions. In section 4 we repeat the 
procedure of section 3 for the case when one or both Majorana masses are
spatially homogeneous. The phenomenological significance of obtained results
is discussed in section 5.

\section{Lagrangian, Ansatz and Equations of Motion}

We consider a Lagrangian in $3+1$ dimensions involving the real scalar 
field $\Phi$ and two chiral spinor fields $\psi_L$ and $\psi_R$ : 

\begin{eqnarray} \label{L}
{\cal L} &=& {1\over 2} \partial_\mu \Phi \partial^\mu \Phi -
{\lambda \over 4} (\Phi^2-\eta^2)^2 \nonumber +\\
&+& i \overline{\psi_L} \gamma^\mu \partial_\mu \psi_L  
+ i \overline{\psi_R} \gamma^\mu \partial_\mu \psi_R  - \\
&-& (k_1 \Phi \overline{\psi_L} \psi_R + k_2 \Phi \overline{\psi_R} \psi_R^c 
+ k_3 \Phi \overline{\psi_L} \psi_L^c + h.c.) \nonumber
\end{eqnarray}
where $\lambda$, $\eta$, $k_1$, $k_2$ 
and $k_3$ are real and positive (if $\psi_L$ and $\psi_R$ are four component
Dirac spinors we can absorb all phases of $k_1$, $k_2$ and $k_3$ into phases
of spinor components). 
$\psi^c \equiv C \bar{\psi}^T$, where $C$ is the charge conjugation 
matrix, is introduced in (\ref{L}) in order to have the most general
mass terms for fermions \cite{GD}. The scalar potential in (\ref{L}) 
has two minima, $\Phi_{min} =\pm \eta$, and in the absence of fermions supports 
the domain wall solution:
\be \label{dw}
\Phi_0=\eta \, \tanh(\frac{\sqrt{\lambda} \eta y}{2})
\ee
for a domain wall located in the $xz$ plane.
The resulting fermionic equations of motion in a domain wall background are:

\begin{eqnarray} \label{eom}
 i\gamma^\mu \partial_\mu \psi_L &=& k_1 \Phi_0 \psi_R + k_3 \Phi_0 \psi_L^c \nonumber \\
 i\gamma^\mu \partial_\mu \psi_R &=& k_1 \Phi_0 \psi_L + k_2 \Phi_0 \psi_R^c
\end{eqnarray}

The representation of the Dirac matrices we will use is:

\begin{eqnarray}
 \gamma^0 &=& \pmatrix{\tau^3&0\cr 0&-\tau^3\cr},\ \ \ \ \
\gamma^1=\pmatrix{i\tau^2&0\cr 0&-i\tau^2\cr}, \nonumber \\
\gamma^2 &=& \pmatrix{-i\tau^1&0\cr 0&i\tau^1\cr}, \ \ \                 
\gamma^3=\pmatrix{0&1\cr -1&0\cr}, \nonumber
\end{eqnarray}
\be
C \equiv i \gamma^2 \gamma^0 = \pmatrix{ 0&-1&0&0 \cr 1&0&0&0 \cr
0&0&0&-1 \cr 0&0&1&0} , \ \  \gamma^5=\pmatrix{0&1\cr 1&0} \ . \nonumber
\ee
where $\tau^i$, $i=1,2,3$ are the Pauli matrices.
In this representation, a four-component Dirac fermion has left and 
right-handed components of the form: 
$\psi^T_L =(\alpha ,\beta , -\alpha , -\beta )$ and 
$\psi^T_R=(\gamma , \delta , \gamma , \delta)$. 

For a domain wall located in the $xz$ plane, $\Phi_0$ is a function of $y$
only and we look for a solution to (\ref{eom}) 
of the form:

\begin{eqnarray} \label{ansatz}
\alpha  &=&  a(y) e^{i\omega t -ik_xx -ik_zz} \nonumber \\
\beta   &=&  b(y) e^{i\omega t -ik_xx -ik_zz} \\
\gamma  &=&  c(y) e^{i\omega t -ik_xx -ik_zz} \nonumber \\
\delta  &=&  d(y) e^{i\omega t -ik_xx -ik_zz} \nonumber
\end{eqnarray}
$a(y)$, $b(y)$, $c(y)$ and $d(y)$ can be taken to be real functions of $y$.
We are interested in zero energy solutions to equation (\ref{eom}) 
such that all spinor components fall off exponentially outside 
the string core (large $y$) and are well-behaved (nonsingular) at the origin 
(small $y$). 

After setting $\omega=k_x=k_z=0$, we get two identical sets of
equations:

\begin{eqnarray} \label{sys1}
a'&=& k_1 \Phi_0 d + k_3 \Phi_0 a  \\
d'&=& k_1 \Phi_0 a + k_2 \Phi_0 d \nonumber 
\end{eqnarray}
\begin{eqnarray} \label{sys2}
b'&=& k_1 \Phi_0 c + k_3 \Phi_0 b \\
c'&=& k_1 \Phi_0 b + k_2 \Phi_0 c  \nonumber 
\end{eqnarray}

\section{Solutions}

Let us first analize the system (\ref{sys1}) for $y \geq 0$ 
($y\leq 0$ is analogous because of the symmetry under $y \rightarrow -y$).
No analytic solution to the system could be found, but we can 
learn something about the structure of solutions by looking at their 
asymptotic behavior. Neglecting a domain wall back-reaction to fermions, 
$\Phi_0 \rightarrow \eta$ at large $y$.
Assuming the asymptotic behavior  $a \sim a_0 e^{s y}$ and $d \sim d_0 e^{t y}$ 
($a_0$, $d_0$, $s$ and $t$ are arbitrary real numbers) we get following conditions:

\be
s = t ,\ \ \
s^2 - s \eta (k_2+k_3) + k_2 k_3 \eta^2 - k_1^2 \eta^2 =0
\ee

The solutions to this second order polynomial equation for $s$ are
\be \label{s}
s_{\pm} \equiv {1 \over 2} \eta \left[(k_2+k_3) \pm \sqrt{(k_2 - k_3)^2 + 
4 k_1^2} \right] 
\ee
Both solutions for $s$ are real. $s_+$ is positive, and is giving rise to exponentially 
growing mode. $s_-$ is negative if and only if $k_2 k_3 < k_1^2$, 
giving rise to exponentially decaying mode. Only
exponentially decaying modes are physically acceptable. So, we conclude that 
the necessary condition for the zero mode to exist on a domain wall is 
$k_2 k_3 < k_1^2$. We should note that this result is very similar to the 
conjecture in \cite{us} about the existence of neutrino zero modes on 
electroweak strings.

The analysis for $y \leq 0$ is analogous. For $y<0$ we have 
$a \sim a_0 e^{-s |y|}$, so only positive
$s$ leads to a physically acceptable result. But we also have 
$\eta \rightarrow -\eta$ in (\ref{s}) and the necessary condition for the
existence of a zero mode stays the same.

If we want to match the exponentially decaying large $y$ solutions in a given
pair of equations to the solutions at the origin, we must know how 
many solutions for small $y$ are well-behaved.
Neglecting a domain wall back reaction to fermions, 
$\Phi_0 \sim  f_0 y + f_2 y^3+ \ldots$  for small $y$. Assuming that
$a \sim a_0 y^s$, $d \sim d_0 y^t$ 
($a_0$ and $b_0$ are arbitrary real numbers while $s$ and $t$
are nonnegative real numbers) for small $y$, we find that the leading orders
are $s=0$, $t=0$. System (\ref{sys1}) is invariant under parity
($y \rightarrow -y$), so all corrections to the solution
which are parity violating are zero. We can write the general well-behaved 
solution in the form:

\be \label{sol}
\left( \begin{array}{c} a \\ b \end{array} \right) 
 = \left( \begin{array}{c}
a_0 y^0 + a_2 y^2 + a_4 y^4 + \ldots \\
d_0 y^0 + d_2 y^2 + d_4 y^4 + \ldots \\ 
\end{array} \right) 
\ee

Substituting (\ref{sol}) into (\ref{sys1}) and equating coefficients of
the same order in $y$ we see that only two coefficients are
independent (say $a_0$ and $d_0$), while 
all others are functions of the first two). This means that there
are two linearly
independent well-behaved solutions. For the system of two linear first order
differential equations we should have two solutions in total, so we conclude
that both solutions are well-behaved. 

\begin{figure}[!ht] 
\epsfxsize = 0.85 \hsize \epsfbox{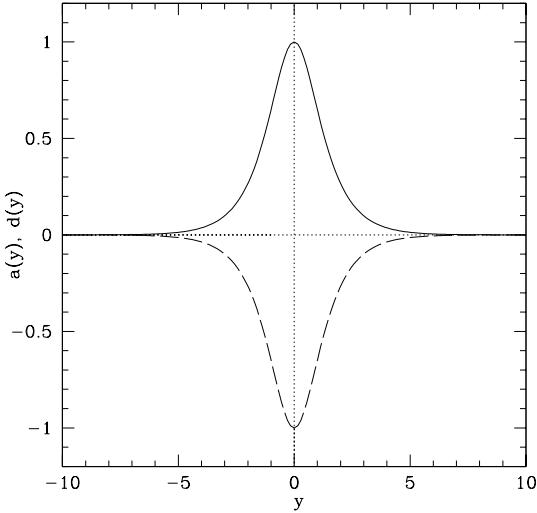} 
\caption{ \label{Fig1} Numerical solution of the system (\ref{sys1}) with 
 $k_1\eta=2$, $k_2\eta=1$, $k_3\eta=1$ and $\frac{\sqrt{\lambda} \eta}{2}=1$, 
 showing $a(y)$ (solid) and $d(y)$ (dashed). The domain wall is located at $y=0$.
 The solution is symmetric with respect to $y \rightarrow -y$.}
\end{figure}

Obviously, all the arguments given for $y \geq 0$ hold for $y\leq 0$ too, so
we will confine our following discussion to positive $y$.
Each of the two well-behaved solutions at the origin matches to
a unique linear combination of both solutions at large $y$. If both solutions
at large $y$ are bad (exponentially growing), then there is no solution
which is well-behaved everywhere. If one of them is good (exponentially
decaying) there is always one well-behaved solution everywhere --- good solution
at large $y$ which matches to a linear combination of two good solutions 
at the origin. In this case, the necessary condition for the zero mode
to exist, $k_2 k_3 < k_1^2$, is also sufficient. 
Indeed, numerically solving the system (\ref{sys1}), we find that there is 
one well-behaved solution (Figure (\ref{Fig1})). 
We took $k_1 \eta =2$, $k_2\eta=1$ and $k_3\eta=1$ which satisfies the necessary and
sufficient condition for a zero mode to exist. We also set 
$\frac{\sqrt{\lambda} \eta}{2}=1$.

The system of equations (\ref{sys2}) have the same structure as the system
(\ref{sys1}) and the previous analysis applies also to (\ref{sys2}).

If the fermions have only Dirac mass terms, i.e. $k_1 \neq 0$, $k_2=0$, $k_3=0$,
it is possible to solve the system (\ref{sys1}) analytically. In this case
we get the well-behaved solution:

\begin{eqnarray} \label{k1}
a(y) &=& A \left[\cosh(\frac{\sqrt{\lambda} \eta y}{2})\right]^{-\frac{2k_1}{\sqrt{\lambda}}} \\
d(y) &=& -A \left[\cosh(\frac{\sqrt{\lambda} \eta
y}{2})\right]^{-\frac{2k_1}{\sqrt{\lambda}}} \nonumber
\end{eqnarray}
where $A$ is a normalization constant. This result is in agreement with the 
results in \cite{Jackiw,Callan}.
Solution (\ref{k1}) suggests that massless fermions, i.e. fermions with 
$k_1 = 0$, $k_2=0$, $k_3=0$, do not have normalizable zero modes. In this case,
the fermion wave function is a constant and the normalization integral diverges.
However, as was shown in \cite{us} for the case of massless neutrino
zero modes on electroweak strings, a correct interpretation of massless
fermion zero modes is within the framework of the continuum.
Keeping $\omega$ in (\ref{ansatz}) we get equations for massless fermions:

\begin{eqnarray}
-\omega a + b' &=& 0 \nonumber \\
 \omega b + a' &=& 0  \\
-\omega c + d' &=& 0 \nonumber \\
 \omega d + c' &=& 0 \nonumber 
\end{eqnarray}
The solutions are plane waves, $a(y)$, $b(y)$, $ c(y)$, $d(y)$ 
$\sim e^{\pm i \omega y}$, which are Dirac delta-function normalizable.
If we look at the massless fermion zero mode ($\omega \rightarrow 0$ limit
of a plane wave solution) as an isolated zero mode, it is
not normalizable. But, this state is actually part of a continuum
spectrum of the theory and is Dirac delta function normalizable.

\section{Constant Majorana Mass Terms}
Probably the most interesting case is when the source of the Majorana masses
$M_L$ and $M_R$ is not the coupling with the field $\Phi$. The Majorana
mass
terms can be taken to be spatially homogeneous and presumed to 
arise from the vacuum expectation value of some field acquired in a phase 
transition above the phase transition of the field $\Phi$. In this case we set 
$k_2 \Phi \equiv M_R$ and $k_3 \Phi \equiv M_L$ in Lagrangian (\ref{L}).

We can carry out a similar analysis to the one done in section 3. 
The equations of motion now are:
%Substituting $k_2 \Phi_0=M_R$ and $k_3 \Phi_0=M_L$ in (\ref{sys1}) and (\ref{sys2})
%we have:

\begin{eqnarray} \label{sys11}
a'&=& k_1 \Phi_0 d + M_L a  \\
d'&=& k_1 \Phi_0 a + M_R d \nonumber 
\end{eqnarray}
\begin{eqnarray} \label{sys22}
b'&=& k_1 \Phi_0 c + M_L b \\
c'&=& k_1 \Phi_0 b + M_R c  \nonumber 
\end{eqnarray}
An analysis of the large $y$
behavior for $y \geq 0$ is analogous to the previous case. Substituting
$k_2\eta \equiv M_R$, $k_3\eta \equiv M_L$ in (\ref{s}) and demanding 
negative solutions for $s$ we find the 
necessary condition for a zero mode to exist: $M_L M_R < M_D^2$, where we
just formally write $k_1\eta=M_D$. For $y \leq 0$, we demand positive
solutions for $s$ and see that if $M_L M_R < M_D^2$ there is  
one decaying mode. 

\begin{figure}[!ht] 
\epsfxsize = 0.85 \hsize \epsfbox{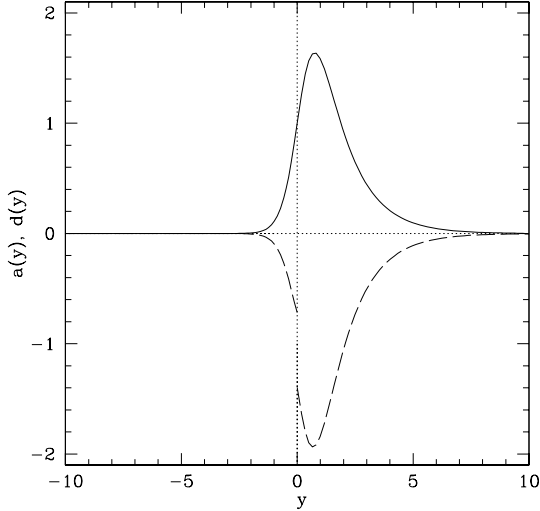} 
\caption{ \label{Fig2} Numerical solution of the system (\ref{sys11}) with 
  $k_1\eta=2$, $M_R=1$, $M_L=1.5$ and $\frac{\sqrt{\lambda} \eta}{2}=1$, 
  showing $a(y)$ (solid) and $d(y)$ (dashed). We see that 
  $\frac{a(y=0+)}{d(y=0+)} \neq \frac{a(y=0-)}{d(y=0-)}$ and the solution
  is not continuous.}
 \end{figure}

\begin{figure}[!ht] 
\epsfxsize = 0.85 \hsize \epsfbox{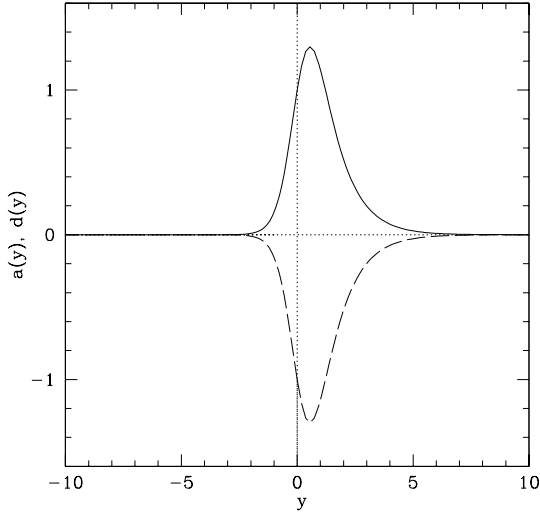} 
\caption{ \label{Fig3} Numerical solution of the system (\ref{sys11}) with 
  $k_1\eta=2$, $M_R=1$, $M_L=1$ and $\frac{\sqrt{\lambda} \eta}{2}=1$, showing 
  $a(y)$ (solid) and $d(y)$ (dashed). The domain wall is located at $y=0$. 
  The solution is not 
 symmetric with respect to $y\rightarrow -y$.}
 \end{figure}

An analysis of the behavior near the origin is slightly different than before 
since the constant $M_L$ and $M_R$ terms break the
$y \rightarrow -y$ symmetry of the equations (\ref{sys11}) and parity 
breaking corrections to the leading order in a solution are also present 
(odd powers in $y$). Thus, the general well-behaved solution to 
the system (\ref{sys11}) can be written as:

\be \label{sol1}
\left( \begin{array}{c} a \\ b \end{array} \right) 
 = \left( \begin{array}{c}
a_0 y^0 + a_1 y + a_2 y^2 + a_3 y^3 + \ldots \\
d_0 y^0 + d_1 y + d_2 y^2 + d_3 y^3 + \ldots \\ 
\end{array} \right) 
\ee
Substituting (\ref{sol1}) into (\ref{sys11}) we see that two coefficients are
independent (say $a_0$ and $d_0$) which means that there are two linearly 
independent well-behaved solutions. This conclusion is valid for both
positive and negative $y$.
The matching of well-behaved solutions at large and small $y$
is similar to system (\ref{sys1}). However, the matching of a given pair of
solutions for $a(y)$ and $d(y)$ for $y \geq 0$ and $y \leq 0$ at $y=0$
is nontrivial. There is no symmetry $y \rightarrow -y$ and 
$\frac{a(y=0+)}{d(y=0+)} \neq \frac{a(y=0-)}{d(y=0-)}$ (Figure (\ref{Fig2})). 
Numerical analysis
shows that $\frac{a(y=0+)}{d(y=0+)} = \frac{a(y=0-)}{d(y=0-)}$ only if $M_L=M_R$.
%Note that in this case the Lagrangian (\ref{L}) is symmetric with respect
%to $\psi_L \rightarrow \psi_R$. 
We conclude that if $M_L M_R < M_D^2$, where $M_L=M_R$, there is one solution
which is well-behaved everywhere. Numerically solving the system (\ref{sys11})
setting $k_1 \eta =2$, $M_R=1$, $M_L=1$ and $\frac{\sqrt{\lambda} \eta}{2}=1$, 
we find one well-behaved solution (Figure (\ref{Fig3})).

Let us mention that a similar analysis can be done in the case when just
one Majorana mass term is homogeneous, while the second one still comes from a
coupling to the field $\Phi$. Setting $k_3 \Phi \equiv M_L$ in (\ref{L})
and repeating
the procedure from section 3 we find the same form of the necessary condition 
for a zero mode to exists $M_L M_R < M_D^2$, where we just formally write 
$k_1\eta=M_D$ and $k_2\eta=M_R$. However, unlike the previous case, we can
not set $M_L = k_2 \Phi_0$ everywhere. Thus, $\frac{a(y=0+)}{d(y=0+)} 
\neq \frac{a(y=0-)}{d(y=0-)}$ for all values of $M_L$ and there is no matching between
the $y \geq 0$ and $y \leq 0$ solutions for $a(y)$ and $d(y)$. In this case
there are no well behaved solutions.

\section{Phenomenology}

By now, we have not addressed the question about the phenomenological
validity of the general fermionic mass terms. The Lagrangian 
(\ref{L}) could be, in principle, part of some larger theory like
the standard model (and various extensions of it) or GUT theories. 
The specific charges assigned to
the fermionic field $\psi$ would then depend on a specific symmetry group
of the model. Mass terms, present in a Lagrangian have to be
gauge singlets, thus restricting their possible forms. An illustrative example, 
phenomenologically valid 
standard model neutrinos, beside the Dirac mass terms, can only
have the right-handed Majorana mass terms. In our notation, this
corresponds to the case of $k_3=0$. If the coupling to the scalar field
$\Phi$ gives rise to the right-handed Majorana mass we analize the results
of section 3. Setting $k_3=0$ in (\ref{s}) we see that the necessary condition
$k_2k_3 < k_1^2$, i.e. $0 < k_1^2$, is always satisfied. Following the rest of
the analysis of section 3, we conclude that in this case there is always one
zero mode. However, in this case it would be difficult, without a fine tunning,
to explain a large difference between the Dirac and right-handed Majorana 
masses because the same phase transition gives the scale for both masses.

If the right-handed Majorana mass is spatially homogeneous, then we apply
the analysis of the section 4 (which is different from the analysis of the
section 3 because there is no symmetry $y \rightarrow -y$). In the absence of 
the left-handed Majorana mass, although the necessary condition is satisfied, 
due to the fact that $0=M_L \neq M_R$, the solutions for positive and negative $y$ 
do not match at $y=0$. In this case, there are no well behaved zero modes.

\section{Conclusion}

We studied zero energy solutions of the Dirac equation in the background of
a domain wall located in the $xz$ plane. We first considered the case when the
vacuum expectation value 
of the scalar field $\Phi$ was the source of the Dirac and Majorana mass terms. 
In the general case, when both the Dirac and Majorana mass terms 
are present we solved the
equations of motion analytically in two asymptotic regimes, large and small $y$ and
found the necessary and sufficient condition for a zero mode to exist.
This condition is very similar to the one conjectured in
\cite{us} in the case of neutrino zero modes on electroweak strings.
We also solved the equations numerically.

If the Majorana mass terms are absent, it is possible to solve the equations
analytically. The result agrees with \cite{Jackiw,Callan}.

If the fermions are massless there exists a zero-energy solution which is
not discretely normalizable. Solving the non-zero energy equations we
showed that this state is actually part of the continuum spectrum and
is Dirac delta function normalizable.

We also considered the cases when one or both Majorana terms are spatially 
homogeneous. If both of them are spatially homogeneous and if $M_L M_R < M_D^2$, 
with $M_L=M_R$, there is one well behaved solution. If just one is 
spatially homogeneous there are no well behaved solutions. The phenomenological
significance of obtained solutions was discussed.

\section{Acknowledgements}

We would like to thank G. Starkman and T. Vachaspati for valuable discussions.
This work was supported in part by DOE.

\end{document}